\documentclass[12pt,nofootinbib,aps,amsmath,amssymb,preprint,showpacs]{revtex4}
\usepackage[T1]{fontenc} \usepackage[latin1]{inputenc}
\usepackage{amsmath,color,ulem} \usepackage{graphicx}
\usepackage{epsf,bm} \usepackage{amssymb}
\begin{document}

\title{Dynamics of a double-stranded DNA segment in a shear flow}

\author{Debabrata Panja} \affiliation{Institute for Theoretical
  Physics, Universiteit Utrecht, Leuvenlaan 4, 3584 CE Utrecht, The
  Netherlands} \author{Gerard T. Barkema} \affiliation{Institute for
  Theoretical Physics, Universiteit Utrecht, Leuvenlaan 4, 3584 CE
  Utrecht, The Netherlands} \affiliation{Instituut-Lorentz,
  Universiteit Leiden, Niels Bohrweg 2, 2333 CA Leiden, The
  Netherlands} \author{J. M. J. van Leeuwen}
\affiliation{Instituut-Lorentz, Universiteit Leiden, Niels Bohrweg 2,
  2333 CA Leiden, The Netherlands}

\begin{abstract} 
  We study the dynamics of a double-stranded DNA (dsDNA) segment, as a
  semiflexible polymer, in a shear flow, the strength of which is
  customarily expressed in terms of the dimensionless Weissenberg
  number Wi. Polymer chains in shear flows are well-known to undergo
  tumbling motion. When the chain lengths are much smaller than the
  persistence length, one expects a (semiflexible) chain to tumble as
  a rigid rod. At low Wi, a polymer segment shorter than the
  persistence length does indeed tumble as a rigid rod. However, for
  higher Wi the chain does not tumble as a rigid rod, even if the
  polymer segment is shorter than the persistence length. In
  particular, from time to time the polymer segment may assume a
  buckled form, a phenomenon commonly known as Euler buckling. Using a
  bead-spring Hamiltonian model for extensible dsDNA fragments, we
  first analyze Euler buckling in terms of the oriented deterministic
  state (ODS), which is obtained as the steady-state solution of the
  dynamical equations by turning off the stochastic (thermal) forces
  at a fixed orientation of the chain. The ODS exhibits symmetry
  breaking at a critical Weissenberg number Wi$_{\text c}$, analogous
  to a pitchfork bifurcation in dynamical systems. We then follow up
  the analysis with simulations and demonstrate symmetry breaking in
  computer experiments, characterized by a unimodal to bimodal
  transformation of the probability distribution of the second Rouse
  mode with increasing Wi. Our simulations reveal that shear can cause
  strong deformation for a chain that is shorter than its persistence
  length, similar to recent experimental observations.
\end{abstract}

\pacs{36.20.-r,64.70.km,82.35.Lr}
\maketitle

\section{Introduction\label{sec1}}

The flow properties of a solution of polymers have attracted the
interest of physicists for a long time. One side of the problem
concerns how the concentration of dissolved polymers influences
e.g. the viscous (or viscoelastic) properties of the fluid. The other
side concerns how fluid flow influences the behavior of polymers. Here
we restrict ourselves to the latter. Specifically, we consider a
dilute solution of double-stranded DNA (dsDNA) segments in water under
shear. Double-stranded DNA is a semiflexible polymer, since it
preserves mechanical rigidity over a range, characterized by the
persistence length $l_p\approx40$ nm \cite{dnapersist,wang}, along its
contour.

That a polymer will go through a ``coil-stretch transition'' under the
influence of a shear flow was originally predicted by de Gennes
\cite{degennes}, although it would be more than two decades before
the coil-stretch transition would be put to experimental
verification. Interestingly however, the first key experiment along
this line --- combining fluid flow and fluorescence microscopy
techniques (the latter in order to visually track polymers) --- was
performed to determine the force-extension curve of dsDNA, wherein
uniform water flow was used to stretch (end-tethered) polymers
\cite{perkins95}. Extending that experimental setup to include more
complicated flow patterns, such as elongational flow
\cite{perkins97,smith98} and shear flow \cite{smith99,leduc99} soon
followed, driven by the quest to understand how flow-induced
conformational changes take place in polymers (see e.g.,
Ref. \cite{shaqfeh} for a review).

An intriguing by-product of the experiments with shear flow was
the tumbling motion of the chains, which can be tracked by, e.g., the
relative orientation of the polymer's end-to-end vector with respect to
the direction of the flow \cite{smith99,leduc99}. Although irregular
at short time-scales, a tumbling frequency could be defined based
on the long-time statistics of the chain's orientation. The tumbling
behavior soon started to receive further attention from researchers:
over the last decade and a half, a number of models have been constructed
\cite{turit1,turit2,lang}
and further experiments have been performed
\cite{doyle,teix,schroeder2,harasim} to characterize and
quantify the tumbling behavior, in particular the dependence of the
tumbling frequency on the shear strength. The subject of this paper,
too, is tumbling behavior in a shear flow, specifically for a dsDNA
chain that is smaller than its persistence length.

As stated earlier, the shear strength $\dot\gamma$ is customarily
expressed by the dimensionless Weissenberg number Wi
$=\dot\gamma\tau$, where $\tau$ is a characteristic time-scale for the
polymer. At one extreme, for flexible polymers (polymer segments that
are many times longer than their persistence length, assuming coil
configurations in the absence of shear), which many of the above
studies focus on, the natural choice for $\tau$ is the polymer's
terminal relaxation time. For them there is good theoretical,
numerical and experimental evidence that the tumbling frequency $f$
scales with Wi as $f\propto$ Wi$^{2/3}$
\cite{doyle,teix,schroeder2,turit1,turit2}. At the other extreme, for
semiflexible polymer segments (polymer segments shorter than their
persistence lengths resemble the configuration of rigid rods in the
absence of shear, and the natural choice of $\tau$ is the time-scale
for rotational diffusion of a rigid rod of the same length), one
expects the rigid rod result, namely that the tumbling frequency
scales as $f\propto$ Wi$^{2/3}$ \cite{jeffery,harasim,bloete}. (Given
that the physics of tumbling is different for flexible and
semiflexible polymers, the similarity in the scaling behavior of $f$
is striking.)

Recently, Harasim {\it et al.\/} \cite{harasim} experimented with
tumbling f-actin segments of several lengths ($\sim$ 3-40 $\mu$m) in a
shear flow. They found that the tumbling frequency $f$ follows the law
$f \propto$ Wi$^{2/3}$ for small Weissenberg numbers. A closer
inspection of their data reveals significant deviations from the $f
\propto$ Wi$^{2/3}$ power-law around and above the persistence length
($\approx$ 16 $\mu$m). Images and movies out of the experiments have
revealed that f-actin segments of lengths smaller than the persistence
length can strikingly buckle into J and U-shapes, broadly known as
Euler buckling.

These issues of buckling and the tumbling frequency were taken up by Lang
et al. \cite{lang} by an extensive modeling study, using the inextensible
wormlike chain as Hamiltonian. They discussed the tumbling frequency
for the whole range spanning the two extremes, i.e., from flexible to
semiflexible polymer segments, and reported, in the intermediate regime,
the dependence $f \propto$ Wi$^{3/4}$.

The present paper has been inspired by the experiment of Harasim {\it
et al}. \cite{harasim}. Our focus is to provide a {\it
quantitative\/} characterization of the Euler buckling, and the
corresponding shapes of a tumbling semiflexible polymer segment in a
shear flow. To this end, we take advantage of a recently developed
bead-spring model for semiflexible polymers \cite{leeuwen,leeuwen1}
and its highly efficient implementation on a computer
\cite{leeuwen2}. We model dsDNA segments dynamics for lengths
$\lesssim20$ nm, and analyze their dynamics in terms of the Rouse
modes \cite{rouse}.  The persistence length of dsDNA is $\approx 40$
nm, corresponding to $\approx 120$ beads with the average intra-bead
distance $\approx0.33$ nm, the length of a dsDNA basepair.  We show
that the tumbling frequency adheres to the rigid rod results
at low Wi and that for high Wi, semiflexible polymer segments tumble much faster. 
This difference quickly leads us to
issues related to (Euler) buckling of the chain under the influence of
shear. We first analyze Euler buckling in terms of the oriented
deterministic state (ODS), which results from turning off the
stochastic (thermal) forces in polymer dynamics at a fixed orientation
of the chain. In this state the internal forces, tending to keep the
chain straight, balance the shear forces.  Below a critical
Weissenberg number Wi$_{\text c}$, the ODS shows a slightly bend
S-shape.  Above Wi$_{\text c}$ a symmetry breaking takes place,
analogous to pitchfork bifurcation, where the ODS strongly deviates
from a rigid rod.

We follow up the ODS analysis with simulations and demonstrate
symmetry breaking in computer experiments, and demonstrate that
similar to the experimental snapshots found for f-actin filaments in
Ref. \cite{harasim}, shear can cause strong deformation, even for a
chain that is shorter than its persistence length.

The structure of the paper is as follows. In Sec. \ref{sec2} we
introduce the model. In Sec. \ref{sec3} we describe the polymer
dynamics in terms of the Rouse modes. In Sec. \ref{sec4} we analyze
the time evolution of the orientation of the polymer, from which we
determine the tumbling frequency. In Sec. \ref{sec5} we analyze Euler
buckling, identify the critical Weissenberg number Wi$_{\text c}$ and
solve for the shapes of the polymer in the ODS. We follow up the
theory of Sec. \ref{sec5} with simulations in Sec. \ref{sec6}, and end
the paper with a discussion in Sec. \ref{sec7}. A movie of a tumbling
dsDNA segment can be found in the ancillary files --- details on the 
movie are provided in Sec. \ref{sec6}.

\section{The model \label{sec2}}

The Hamiltonian for our bead-spring model for semiflexible polymers,
the details of which can be found in our earlier works
\cite{leeuwen,leeuwen1,leeuwen2}, reads
\begin{equation} 
  {\cal H} =\frac{\lambda}{2} \sum^N_{n=1} (|{\bf u}_n|-d)^2 - 
  \kappa \sum^{N-1}_{n=1} {\bf u}_n \cdot {\bf u}_{n+1},
\label{a1}
\end{equation}  
with stretching and bending parameters $\lambda$ and $\kappa$
respectively.  Here ${\bf u}_n$ is the bond vector between the
$(n-1)$-th and the $n$-th beads
\begin{equation} 
  {\bf u}_n = {\bf r}_n - {\bf r}_{n-1},
\label{a2}
\end{equation} 
and ${\bf r}_n$ is the position of the $n$-th bead
($n=0,1,\ldots,N$). The parameter $d$ provides a length-scale, by
the use of which we reduce the Hamiltonian to
\begin{eqnarray}
  \frac{\cal
    H}{k_BT}\!=\!\frac{1}{T^*}\!\left[\sum^N_{n=1} (|{\bf
      u}_n|\!-\!1)^2\!-\!  2 \nu\!\! \sum^{N-1}_{n=1}\! {\bf u}_n\!  \cdot\!
    {\bf u}_{n+1}\! \right]\!\!,
\label{e02}
\end{eqnarray} 
with dimensionless $\nu=\kappa/\lambda$ and $T^*=k_BT/(\lambda d^2)$
parametrizing the Hamiltonian. In this formulation the persistence
length of the polymer is given by $l_p=(\nu/T^*)d/(1-2\nu)$. The model
is a discrete version of the polymer with $N$ discretization units
(i.e., of length $N$). From the analysis of the ground-state of the
Hamiltonian (\ref{e02}) \cite{leeuwen,leeuwen1,leeuwen2}, each
discretization unit can be shown to have a length $a=d/(1-2\nu)$.

The parameters of the model --- $T^*$ and $\nu$ --- are determined by
matching to the force-extension curve. For dsDNA, our semiflexible
polymer of choice in this paper, we use $a=0.33$ nm, the length of a
dsDNA basepair, which leads to $T^*=0.034$ and $\nu=0.353$, meaning
that one persistence length corresponds to $N\approx 120$
\cite{leeuwen,leeuwen1,leeuwen2}.

\section{Polymer dynamics\label{sec3}}

\subsection{Construction of the Rouse modes
  modes\label{sec3a}}

We analyze the dynamics of the polymer by its Rouse modes, since they
turn out to be a convenient scheme for solving the equations of motion
with a sizable time step, without introducing large errors
\cite{leeuwen2}.

The representation of the configurations of a polymer chain in terms
of its fluctuation modes uses basis functions. The well-known Rouse
modes employ the basis functions
\begin{equation} 
  \phi_{n,p}= \left(\frac{2}{N+1}\right)^{1/2} \cos
  \left[\frac{(n+1/2)p\pi}{N+1} \right],
\label{b1}
\end{equation} 
such that conversion of positions ${\bf r}_n$ to Rouse modes ${\bf
  R}_p$ and vice versa given by
\begin{equation} 
  {\bf R}_p = \sum_n {\bf r}_n \phi_{n,p}, \quad \quad
  \quad {\bf r}_n = \sum_p \phi_{n,p} {\bf R}_p,
\label{b2}
\end{equation} 

The modes with $p=0$ correspond to the location of the center-of-mass,
the dynamics of which can be rigorously separated from that of the
other modes. We eliminate the center-of-mass motion by always
measuring the bead positions with respect to the center-of-mass.

\subsection{The equations for the Rouse modes under
  shear\label{sec3b}}

We consider the situation where water flows in the $\hat{\bf
  x}$-direction, with a shear gradient $\dot\gamma$ in the $\hat{\bf
  y}$-direction. The Langevin equation for the motion of the bead
position ${\bf r}_n$ then reads
\begin{equation} 
  \frac{d {\bf r}_n}{ dt} = -\frac{1}{\xi}\frac{\partial{\cal H}}{\partial 
{\bf r}_n} +  \dot{\gamma} \, (y_n -Y_{cm}) \, \hat{\bf x} +\bm{k}_n.
\label{c1}
\end{equation}
The Hamiltonian ${\cal H}$ is given in Eq. (\ref{a1}), and $\xi$ is
the friction coefficient due to the viscous drag, acting on each
bead. The first term on the right hand side of the equation represents
the internal force, which tends to keep the chain straight.  The
second term is the shear force due to the flow, where $\dot{\gamma}$
is the shear rate, $Y_{cm}$ is the $y$ co-ordinate of the
center-of-mass of the chain and $y_n$ is the $y$ co-ordinate of monomer
$n$. As mentioned earlier, we measure the bead positions wrt the
location of the chain's center-of mass, leading to the term $\propto
(y_n -Y_{cm})$. The last term in Eq.  (\ref{c1}) gives the influence
of the random thermal force $\bm{k}_n$, which has the correlation
function
\begin{equation} 
  \langle k^\alpha_n (t) \, k^\beta_m (t' ) \rangle = (2 \, k_B \, T /\xi)  \, 
  \delta^{\alpha,\beta} \, \delta_{n,m} \delta (t-t').
\label{c2}
\end{equation} 

In order to work with dimensionless units we replace the time $t$ by
\begin{equation} 
  \tau =\lambda t / \xi.
\label{c3}
\end{equation}
The ratio $\xi/ \lambda$ then becomes the microscopic time scale, such
that $\tau$ is dimensionless.  In the same spirit we combine the shear
ratio $\dot{\gamma}$ with this time scale leading to the dimensionless
constant $g$ as the shear strength
\begin{equation} 
  g = \dot{\gamma} \, \frac{\xi}{\lambda}.
\label{c4}
\end{equation} 
The shear strength is customarily expressed in terms of the Weissenberg
number, which we define as
\begin{equation} 
  \mbox{Wi} = \frac{\dot{\gamma}}{2 D_r} \quad \quad \quad {\rm with} \quad \quad 
  D_r = \frac{k_B T}{I \xi}.
\label{c5}
\end{equation} 
Here $D_r$ is the rotational diffusion constant with $I$ as the moment
of inertia of the polymer segment in its ground-state (of the
Hamiltonian). The relation between the two dimensionless quantities Wi
and $g$ is then given by
\begin{equation} \label{cn}
\mbox{Wi} = g \frac{I_0}{2 T^*}
\end{equation} 
where $I_0=I/d^2$, the dimensionless moment of inertia of the polymer
segment in the ground-state.

Using the orthogonal transformation converting positions into modes
the dynamic equations for the Rouse modes can be cast in the form
\cite{leeuwen2}
\begin{equation} 
  \frac{d {\bf R}_p}{ d \tau} =-\zeta_p {\bf R}_p +{\bf F}_p+{\bf H}_p +
  \bm{K}_p.
\label{c6}
\end{equation}
For the decay constant we use the expression
\begin{equation} 
  \zeta_p = 4 \nu \left[1 - \cos \left(\frac{p \pi}{N+1}\right)\right]^2.
\label{c7}
\end{equation} 
This spectrum follows from a subtraction in the coupling
force ${\bf H}_p$, which derives from the contour length term in the
Hamiltonian \cite{leeuwen2}
\begin{equation} 
  {\bf H}_p = \left(\frac{2}{N+1}\right)^{1/2} \sum_n
  \sin \left( \frac{p n \pi}{N+1} \right) {\bf u}_n \left(
    \frac{1}{u_n} - 1+2 \nu\right).
\label{c8}
\end{equation}
The subtraction $1- 2\nu$ within the last brackets changes the Rouse
spectrum from longitudinal to the transverse form Eq. (\ref{c7}).
Finally, ${\bf F}_p$ is the shear force given by
\begin{equation} 
  {\bf F}_p = g \, \hat{\bf x} \, (\hat{\bf y} \cdot
  {\bf R}_p).
\label{c9}
\end{equation} 
The fluctuating thermal force $K^\alpha_p$ is the orthogonal
transform of the $\bm{k}_n$ in Eq. (\ref{c2})
\begin{equation} 
  \langle K^\alpha_p (\tau) K^\beta_q (\tau')  \rangle = 2 T^* \delta^{\alpha, \beta} \, 
  \delta_{p,q} \, \delta (\tau - \tau').
\label{c10}
\end{equation} 

Although the Rouse modes turn out to be a convenient scheme for
solving the equations of motion with a sizable time step without
large errors, we do pay a computational penalty in the
calculation of the coupling force, which requires a transformation
(\ref{c3}) from the Rouse modes to the bond vectors and the
transformation (\ref{c8}) back to the modes. The penalty can be kept
to the minimum by the use the fast Fourier transform (FFT) to switch
from modes to bead positions; it keeps the number of operations of the
order $N \log N $.

\subsection{Body-fixed co-ordinate system to analyze tumbling
  dynamics\label{sec3c}}  

One of the major quantities of interest in the tumbling process is the
dynamics of the orientation of the polymer. The orientation can be
defined in several ways. The most common one is the direction of the
end-to-end vector. Since the ends of the chain fluctuate substantially
over short time-scales, this is not a slow variable. We prefer to use
as orientation the direction of the first Rouse mode ${\bf R}_1$,
being the slowest decaying mode. We therefore define the orientation
$\hat{\bf n}$ of the polymer as
\begin{equation} 
  \hat{\bf n} = \hat{\bf R}_1.
\label{d1}
\end{equation} 
We refer to the components of the Rouse modes in the direction of
$\hat{\bf n}$ as longitudinal components
\begin{equation} 
  R^l_p = \hat{\bf n} \cdot {\bf R}_p.
\label{d2}
\end{equation} 
The perpendicular directions are transverse to $\hat{\bf n}$. The
first Rouse mode has, by definition, only a longitudinal component.
In practice the so-defined orientation does not differ much from the 
direction of the end-to-end vector.

Further, it is convenient to discuss the temporal behavior of the
polymer not only in the lab-frame co-ordinate system, with co-ordinate
axes $(\hat{\bf x}, \hat{\bf y},\hat{\bf z})$, but also in the
body-fixed co-ordinate system which we define as follows. Along with
the unit vector $\hat{\bf n}$, one of the two transverse axes 
$\hat{\bf  n}$, is taken perpendicular to $\hat{\bf n}$ and $\hat{\bf
  x}$, namely 
\begin{equation} 
  \hat{\bf m}= \hat{\bf n} \times \hat{\bf x}/r,  \quad \quad  r=[n^2_y
+n^2_z]^{1/2}.
\label{d3}
\end{equation} 
The mode component in this direction, being perpendicular to $\hat{\bf
  x}$, is not influenced by the shear force. The other transverse
direction is then naturally obtained as
\begin{equation} 
  \hat{\bf s} =\hat{\bf m} \times  \hat{\bf n}.
\label{d4}
\end{equation} 
Vector components along $\hat{\bf s}$ are maximally sheared. The
system $(\hat{\bf n}, \hat{\bf s}, \hat{\bf m})$ forms an orthogonal
basis set. For later use, below we list the Cartesian components of
the vectors $(\hat{\bf n}, \hat{\bf s}, \hat{\bf m})$:
\begin{equation} 
  \left\{ \begin{array}{lll}
      n_x = \sin \theta \cos \phi, \quad & s_x=r,  & m_x = 0,  \\*[2mm]
      n_y = \sin \theta \sin \phi, & s_y = - n_x n_y/r, \quad & m_y =  n_z /r, \\*[2mm]
      n_z = \cos \theta, & s_z = - n_x n_z /r, & m_z = - n_y /r. 
\end{array} \right.
\label{d5}
\end{equation} 
We denote the components of the modes generically with the index
$\alpha$, which alternatively runs through $\alpha=(x,y,z)$ or
$\alpha=(n,s,m)$.

\section{Time evolution of the orientation of the polymer\label{sec4}}

The evolution of the orientation is given by the dynamics of the two
transverse components of ${\bf R}_1$
\begin{equation} 
  \frac{d \, \hat{\bf n}}{d \tau} =\frac{d}{d \tau} \frac{{\bf R}_1}{R^l_1} =
  \frac{ d {\bf R}_1}{d \tau} \frac{1}{R^l_1} + {\bf R}_1 \frac{ d }{d \tau} \frac{1}{R^1_1} =
  \frac{1} {R^l_1} \frac{d \, {\bf R}_1}{d \tau},
\label{e1}
\end{equation} 
wherein the third equality follows from the fact that the transverse
components of ${\bf R}_1$ vanish by definition. So the temporal
derivative of the longitudinal component is multiplied the vanishing
transverse component. We then use Eq. (\ref{c6}) and get
\begin{equation} 
  \frac{d \, \hat{\bf n}}{d \tau} = g \, \hat{\bf x} \, (\hat{\bf y} \cdot \hat{\bf n})
  + ({\bf H}_1 + \bm{K}_1)/R^l_1.
\label{e2}
\end{equation}  
Obviously, in the right hand side of the equations only the transverse
components of the vectors are relevant.

For the interpretation of Eq. (\ref{e2}) we note that $R^l_1$ is
closely related to the moment of inertia $I$ of the chain, which is
defined as
\begin{equation} 
  \frac{I}{d^2} = \sum_n (r^l_n)^2 = \sum_p (R^l_p)^2 = \sum_p I_p,
\label{e3}
\end{equation} 
where $I_p$ is the contribution of the $p$-th Rouse mode to the moment of inertia.
When the chain is (relatively) straight, the sum over the modes is
heavily dominated by the first component $I_1$.  So it is an
indicative approximation to replace $I$ by $I_1$.

In the (relatively) straight state the configuration of the chain
resembles that of a straight rod. In order to make a connection with
the equation of tumbling for a rigid rod, we rewrite the equation
using a different scaling of the time $\tau$. We define the variable
$\tilde{\tau}$, linked to the Weissenberg number, as
\begin{equation} 
  \tilde{\tau} = g \tau /{\text{Wi}}, \quad \quad {\rm with} \quad \quad g/{\text{Wi}}=  2 T^*/I_1,
\label{e4}
\end{equation} 
where we have used $I_1$ as measure for the moment of inertia.
In terms of $\tilde{\tau}$, Eq. (\ref{e2}) then becomes
\begin{equation} 
  \frac{d \, \hat{\bf n}}{d \tilde{\tau}} ={\text{Wi}}\,\,[\hat{\bf x} \, (\hat{\bf y}
  \cdot \hat{\bf n})]+ (\tilde{\bf H}_1 + \tilde{\bm{K}}_1), 
\label{e5}
\end{equation} 
with $\tilde{\bf H}_1$ and $\tilde{\bf G}_1 $ defined as
\begin{equation} 
  \tilde{\bf H}_1 = \frac{\sqrt{I_1}}{2 T^*} \,
  {\bf H}_1, \quad \quad \quad \tilde{\bm{K}}_1 (\tilde{\tau}) =
  \frac{\sqrt{I_1}}{2 T^*} \, \bm{K}_1 (\tau).
\label{e6} 
\end{equation}   
The new random force has a correlation function 
\begin{equation} 
  \langle \tilde{K}^\alpha_1 (\tilde{\tau}) \tilde{K}^\beta_1 (\tilde{\tau}') 
\rangle   = \delta^{\alpha, \beta} \delta (\tilde{\tau} - \tilde{\tau}').
\label{e7}
\end{equation}    

Apart from the mode-coupling term $\tilde{\bf H}_1$, Eq. (\ref{e5}) is
the same as that of an infinitely thin rigid rod
\cite{bloete}. Therefore it makes sense to compare the tumbling
frequency $f$ with that of the rigid rod, given by \cite{bloete}
\begin{equation} 
  f = \frac{{\text{Wi}}}{4 \pi (1 +0.65{\text{Wi}}^2)^{1/6}}.
\label{e8}
\end{equation} 

\begin{figure}[h]
\begin{center}
\includegraphics[width=0.55\linewidth]{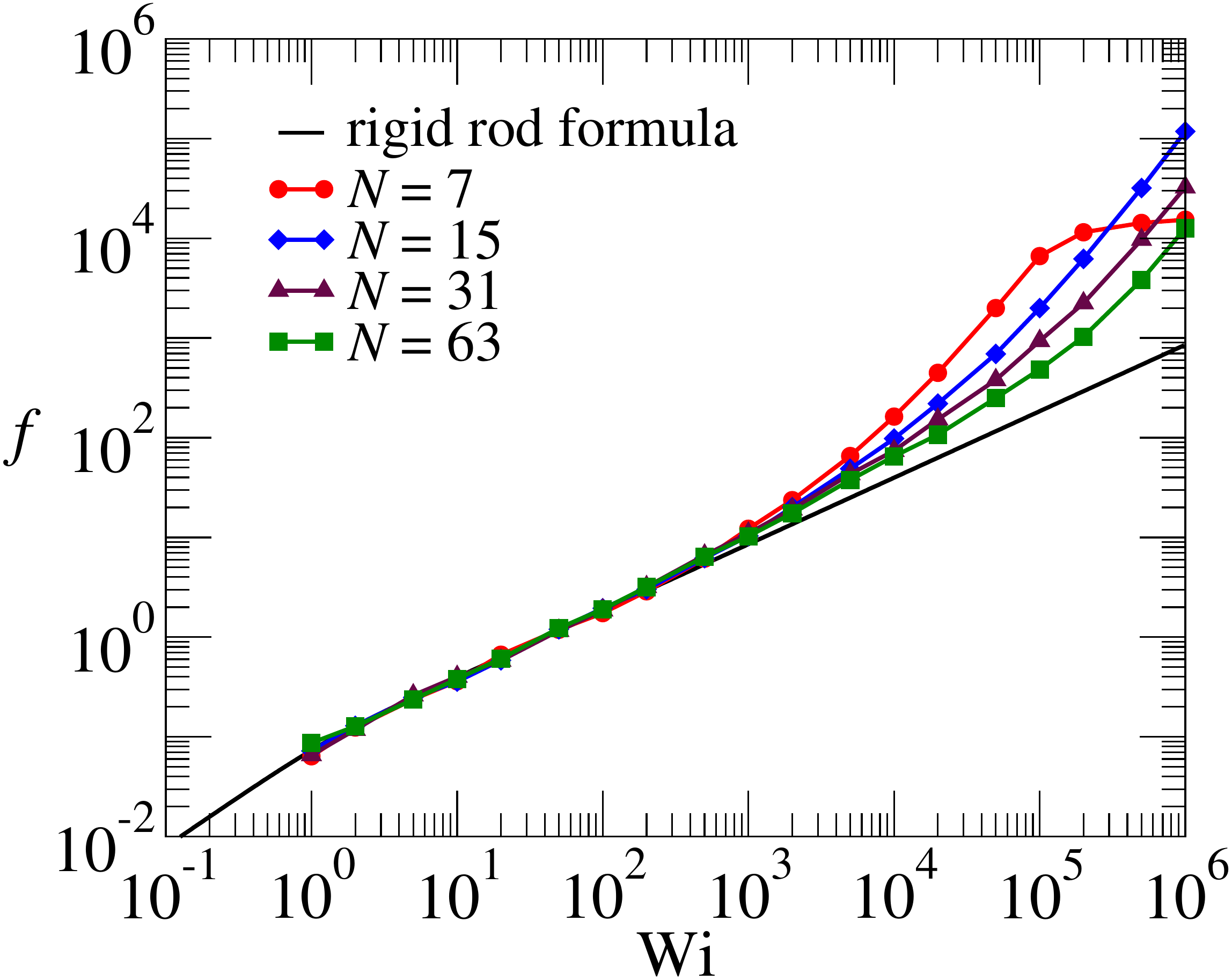}
\end{center}
\caption{(color online) The tumbling frequency as a function of the
  Weissenberg number Wi for a series of chain lengths $N=7,15,31$ and
  63. The conversion to Weissenberg numbers is based on the
  ground-state moment of inertia, see Eq. (\ref{cn}).\label{raw}}
\end{figure}

For comparison we show in Fig. \ref{raw} the tumbling frequency of
dsDNA chains for several lengths shorter than the persistence length
(which corresponds to $N \approx 120$), as found from simulations,
together with that of a rigid rod expression, Eq.~(\ref{e8}). The
simulations have been performed using an efficient implementation of
semiflexible polymer dynamics \cite{leeuwen2} of the bead-spring model
\cite{leeuwen,leeuwen1}. At any given value of the Weissenberg number,
obtained by using the rotational inertia of a rigid rod that has the
same configuration as the ground-state of the Hamiltonian (\ref{e02}),
snapshots of the polymer have been used to calculate its orientation
$[\theta(t),\phi(t)]$ in the laboratory frame. The $\phi(t)$ data is
then fitted by a straight line to obtain the tumbling frequency.

We point out that the simulations follow the rigid rod formula for a
surprisingly large range of Weissenberg numbers, clearly indicating
that up to Wi = 100 the mode-coupling force $\tilde{\bf H}_1$ is
unimportant.  In order to see what this implies for the shear rate
$\dot{\gamma}$, using the expression $I/d^2=N^3/12$ for the moment of
inertia, we write the relation between Wi and the shear rate
$\dot{\gamma}$ as
\begin{equation} 
{\rm Wi} = \dot{\gamma} \frac{a^2 \xi}{k_B T} \frac{N^3}{24}.
\label{e9}
\end{equation}
Note that in Eq. (\ref{e9}) the molecular time scale equals \cite{leeuwen2}
\begin{equation} 
\frac{a^2 \xi}{k_B T} = 52 \times 10^{-12}\, {\rm s}.
\label{e10}
\end{equation} 

Commercially available rheometers at present are limited to shear
rates $\dot{\gamma} < 10^6$ s$^{-1}$. This implies, for dsDNA
fragments of the order of the persistence length, say $N=100$, that
only the range Wi $< 2$ is presently achievable in the lab; i.e., the
differences from the rigid rod behavior in Fig. \ref{raw} lie outside
the reach of present day experiments. Nevertheless, the origin of the
deviations from the rigid rod behavior is theoretically interesting;
we will address this issue in the Sec. \ref{sec6}.

\section{Shapes of semiflexible polymer in the oriented 
deterministic state\label{sec5}} 

In order to further analyze the tumbling process, it is useful to note
that the orientation changes at a slower rate than all the other
modes. This prompts us to focus on the configuration which is obtained
as the steady- state solution of the dynamical equations by turning
off the stochastic (thermal) forces at a fixed orientation of the
chain. We call this configuration the oriented deterministic state
(ODS). We use the properties of the ODS as indicative for the
configurations of the chain at the given orientation.

\subsection{The approach to the oriented deterministic state (ODS) \label{sec5a}}

The  ODS configuration of the chain is obtained from Eq. (\ref{c6}) 
by the decay of the equation
\begin{equation} 
  \frac{d {\bf R}_p}{ d \tau}  = -\zeta_p {\bf R}_p +{\bf F}_p+{\bf
    H}_p.
\label{f1}
\end{equation} 
The constraint of a fixed orientation is imposed by leaving out the
transverse components of the mode ${\bf R}_1$ and setting them equal
to zero in the other mode equations.  Asymptotically the configuration
obeying Eq. (\ref{f1}) will turn into the ODS. So for the ODS the
l.h.s. of Eq. (\ref{f1}) vanishes.  The approach to the ODS
configuration as following from Eq. (\ref{f1}) is slow.

A further simplification of finding the asymptotic state of
Eq. (\ref{f1}) follows by considering the ODS in the body-fixed
system. As there are no shear forces in the $\hat{\bf m}$ direction,
the ODS shape has no component in that direction. For the two other
equations in the $(\hat{\bf n}, \hat{\bf s})$ plane we get in detail
\begin{equation} 
  \left\{ \begin{array}{rcl}
      (\zeta_p - g n_x n_y) R^n_p - g n_x s_y R^s_p & = & H^n_p \\*[4mm]
      g s_x n_y R^n_p + (\zeta_p - g s_x s_y) R^s_p & = & H^s_p
\end{array} \right.
\label{f2}
\end{equation} 
For $p=1$ we have only the first equation since the second refers to
the transverse component $R^s_1$, which we keep equal to zero.
Solving this set of non-linear equations is delicate. We found that,
under normal circumstances, iteration is a stable and quick way to the
solution. For a given orientation of the chain, we start with an
arbitrary configuration (in fact, for the starting configuration, we
use the ground-state configuration of the chain
\cite{leeuwen,leeuwen1}). We then compute the coupling forces $H^n_p$
and $H^s_p$, solve the two-by-two equations (\ref{f2}) for $R^n_p$ and
$R^s_p$ and construct a new set of bond vectors. We repeat the
calculation of $H^n_p$ and $H^s_p$ for the new configuration and
continue the process until the iterative process converges.  Iteration
leads faster to the ODS than the evolution of the equations
Eq. (\ref{f1}). The results of the two approaches, in any case,
coincide.

\subsection{Symmetry breaking in the oriented deterministic
  state \label{sec5b}}

The iterative solution of Eq.~(\ref{f2}), as well as the decay towards
the ODS on the basis of Eq.~(\ref{f1}), reveals an interesting
phenomenon. To show this, we note that configurations that are
invariant under reversal of the chain have vanishing even Rouse
modes. It is easy to see that the equations (\ref{f2}) preserve this
symmetry under iteration. The bond vectors ${\bf u}_n$ changes sign
under the operation
\begin{equation} 
  n  \leftrightarrow N-n.
\label{g1}
\end{equation}
Changing the summation variable from $n$ to $N-n$ in the definition
Eq.  (\ref{c8}) of the coupling force shows that ${\bf H}_p$ changes
sign for even $p$, but not for odd $p$. This means that if we start
the iteration with a configuration that is invariant under reversal,
i.e., we start the iteration with only odd ${\bf H}_p$ on the rhs of
Eq.~(\ref{f2}), it leads to a solution that has, once again, only odd
Rouse mode components.
\begin{figure}[h]
\begin{center}
  \includegraphics[width=0.55\linewidth]{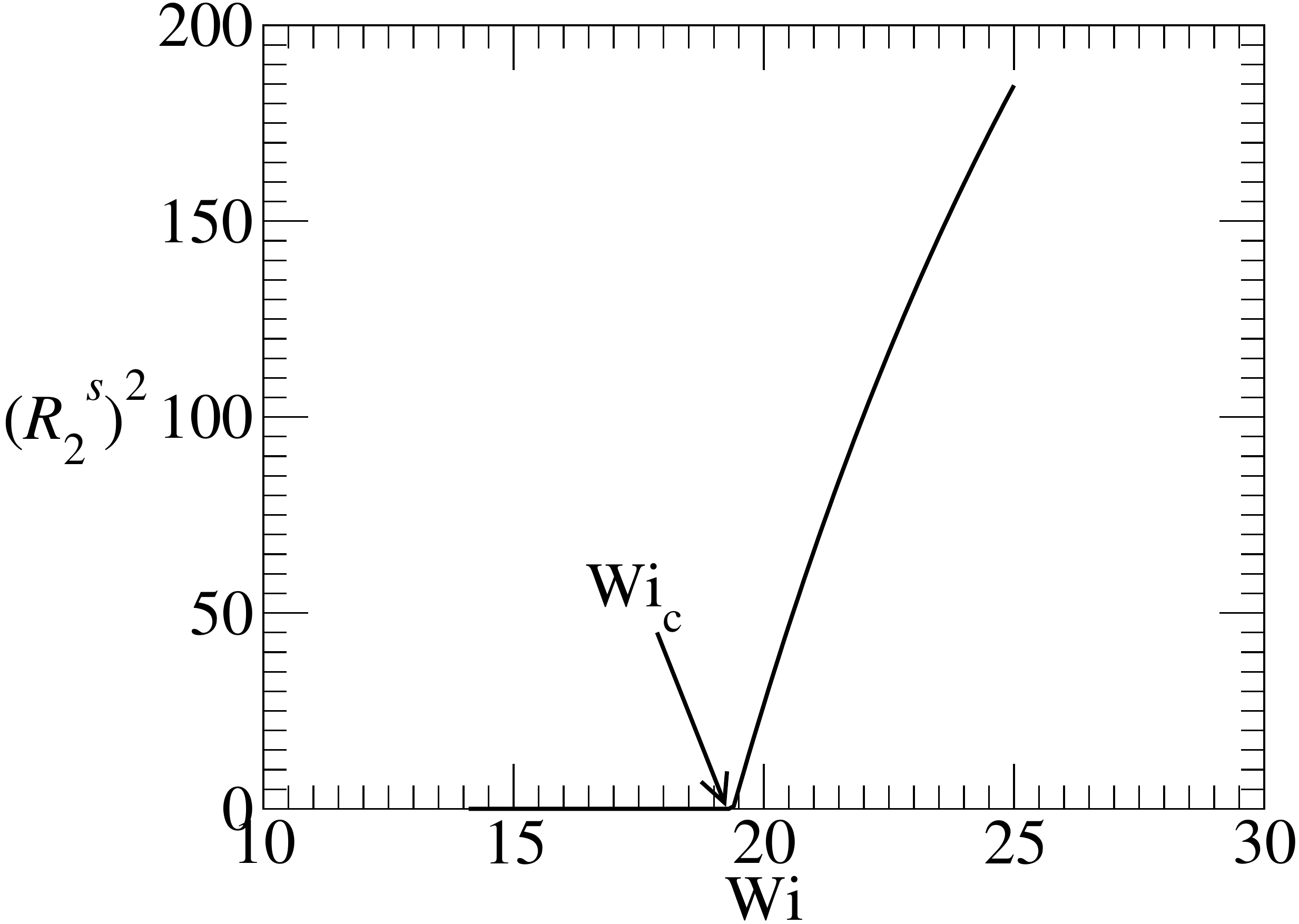}
\end{center}
\caption{The squared value of the transverse component $R^s_2$ as
  function of the Weissenberg number for a dsDNA chain of length
  $N=63$. The critical Weissenberg number is Wi$_{\text
    c}\approx19.4$.}
\label{Weiss}
\end{figure}
\begin{figure*}
\begin{center}
\includegraphics[width=0.465\linewidth]{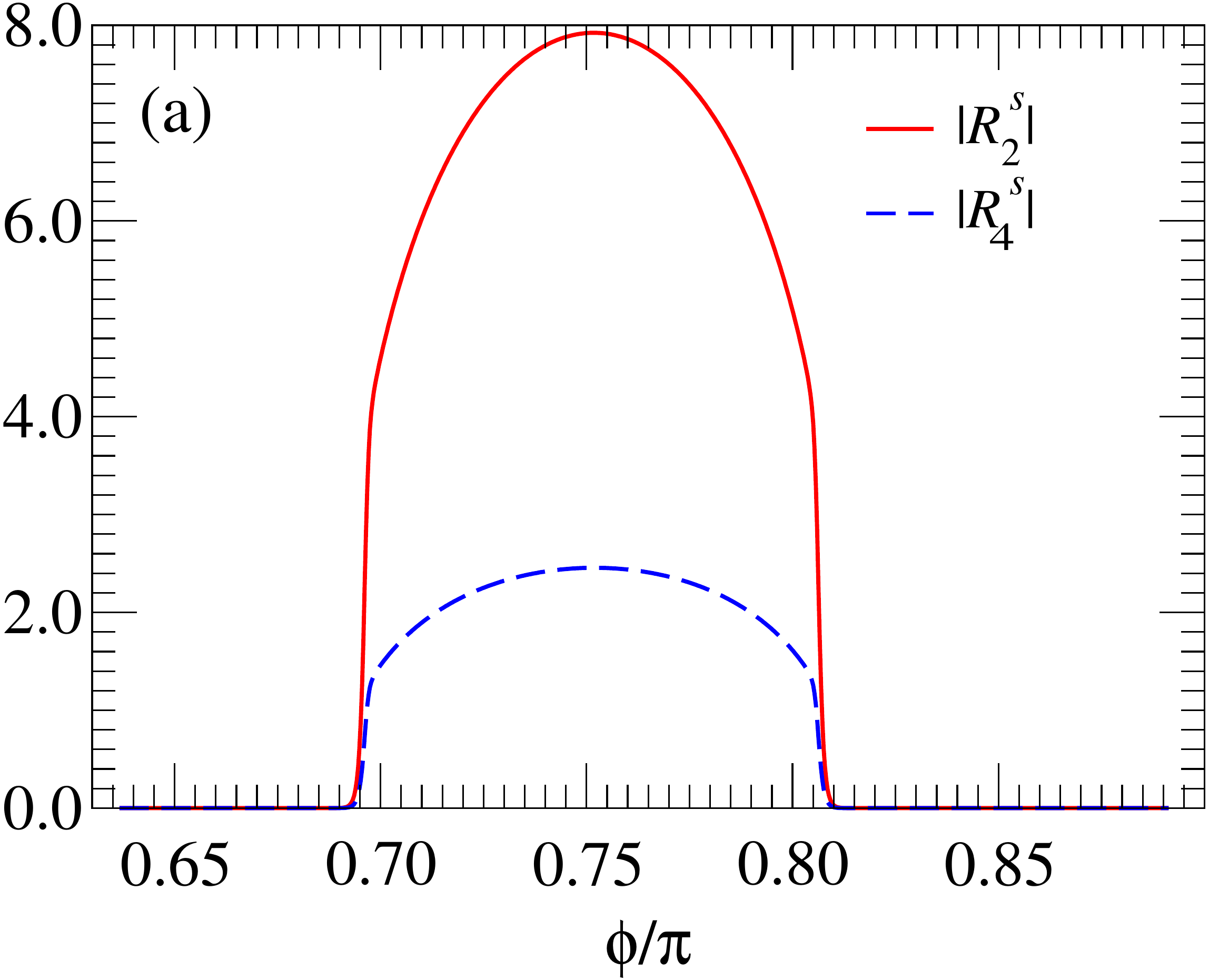}
\hspace{5mm}
\includegraphics[width=0.475\linewidth]{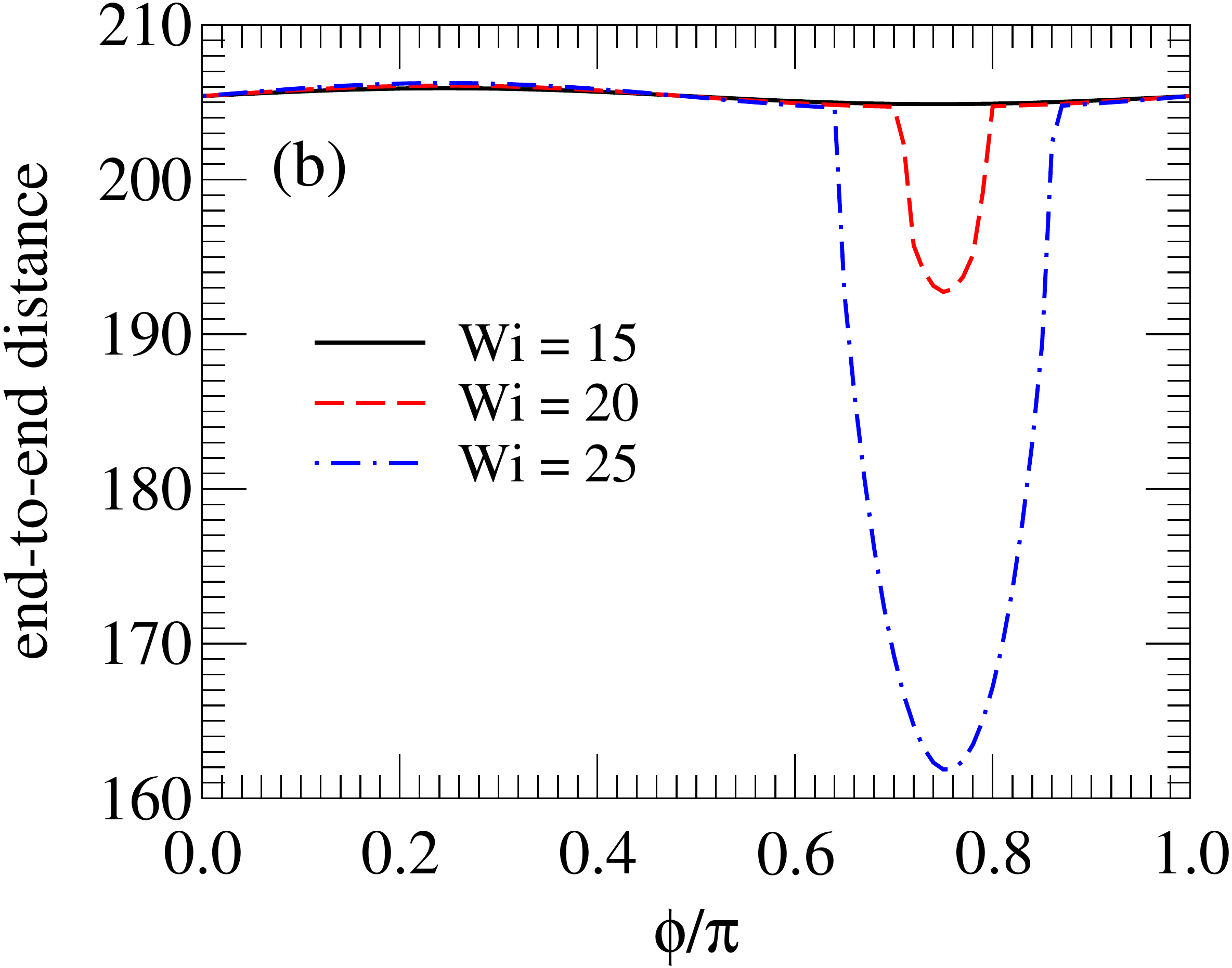}
\end{center}
\caption{(color online) (a) The appearance of the first two even Rouse
  modes in a window around the optimal value $\phi = 3 \pi/4$ for
  $\theta=\pi/2$ at Wi$=21.3$. (b) The end-to-end distance as function
  of $\phi$ for $\theta=\pi/2$ for some values of Wi around Wi$_{\text
    c}\approx19.4$.}
\label{equator}
\end{figure*}

The above does not however exclude that there are solutions which
break the reversal symmetry. The best way to solve for the ODS is to
therefore start the iteration with a configuration with a
(perturbatively) small even mode, e.g. ${\bf R}_2$. The perturbation
may grow or decrease under successive iterations. We find that for low
Weissenberg numbers the perturbation decays to zero, while beyond a
critical Weissenberg number Wi$_{\text c}$, the reversal symmetry is
broken, i.e., the perturbation grows and saturates at a non-zero
value, much like the classic case of a pitchfork bifurcation.

As an example, for a dsDNA chain of length $N=63$ (note: a dsDNA
segment of one persistence length corresponds to $N\approx120$), we
plot the squared value of the transverse component $R^s_2$ as function
of the Weissenberg number in Fig. \ref{Weiss}. At Wi $=$ Wi$_{\text
  c}$ the first non-zero even Rouse modes in the chain appear for
$\theta=\pi/2$ and $\phi=3 \pi/4$. The coefficient of $R^s_2$ in
Eq. (\ref{f2})
\begin{equation} 
  \zeta_p - g s_x s_y=\zeta_p + g (\sin \theta)^2 \sin \phi \cos \phi
\label{g2}
\end{equation} 
reaches its smallest value for $\theta=\pi/2$ and $\phi=3 \pi/4$,
thus leading to the largest value of $R^s_2$ in the case of symmetry
breaking.

From Fig. \ref{Weiss} we see that the critical Weissenberg number for
$\theta=\pi/2$ and $\phi=3 \pi/4$ equals Wi$_{\text c}\approx19.4$ for
a dsDNA chain of length $N=63$.

\subsection{Shapes of the chain and Euler buckling\label{sec5c}}

The shape of the chain depends on its orientation of the polymer,
which comes into the solution through the components of the axes
$\hat{\bf n}$ and $\hat{\bf s}$. The shear is most effective in the
$x$-$y$ plane, i.e., for $\theta=\pi/2$. In Fig. \ref{equator}(a), for
$N=63$, we show the value of $|R^s_2|$ and $|R^s_4|$ as a function of
$\phi$ in the neighborhood of the most effective value $\phi=3 \pi/4$
for Wi$=21.3$ and $\theta=\pi/2$ (note: Wi$_{\text c}\approx19.4$). The
non-zero value of $|R^s_p|$ disappears at $\phi=3 \pi/4$  when Wi
approaches Wi$_{\text c}$ from above.

Further, in order to see the magnitude of the effect we plot in
Fig. \ref{equator}(b) the behavior of the end-to-end distance of the
chain for $N=63$ as a function of $\phi$ for $\theta=\pi/2$ and for
some values of Wi around the critical Weissenberg number Wi$_{\text
  c}$. One observes that the end-to-end distance varies only slightly
as a function of orientation below Wi$_{\text c}$. Above the
Wi$_{\text c}$ a large dip develops around $\phi=3 \pi/4$,
demonstrating that the symmetry breaking goes hand-in-hand with the
so-called Euler buckling of the chain, i.e., the chain folds, which
reduces its end-to-end distance.

\begin{figure}[h]
\begin{center}
\includegraphics[width=0.55\linewidth]{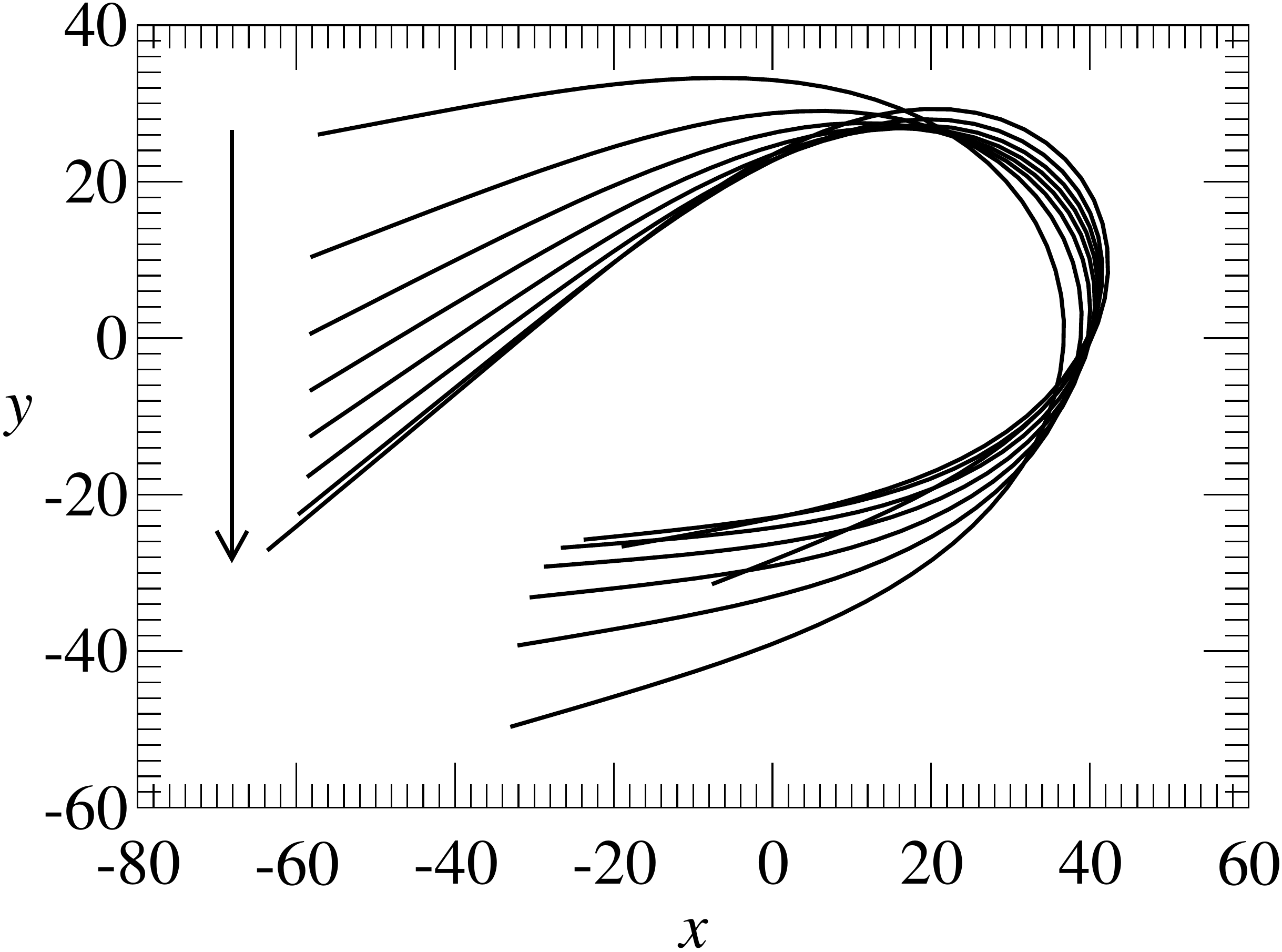}
\end{center}
\caption{Shapes of the dsDNA chain of length $N=63$ at Wi$=100$,
  demonstrating Euler buckling in the ODS. The shapes are shown on the
  $x$-$y$ plane (i.e., $\theta=\pi/2$) for $0.6\pi\leq\phi<\pi$. In
  the direction of the arrow the shapes correspond to $\phi=0.60\pi,
  0.65\pi, 0.70\pi, 0.75\pi, 0.80\pi, 0.85\pi, 0.90\pi$ and $0.95\pi$
  respectively. Interestingly, the orientation here defined by the first Rouse
  mode corresponds closely to the direction of the end-to-end vector.}
\label{shapes}
\end{figure}

In order to visually appeal the reader to Euler buckling, we
provide a number of snapshots of the chain in the ODS for $N=63$ and
Wi $=100$, confined to the $x$-$y$ plane in Fig. \ref{shapes}. This
large Weissenberg number is well above Wi$_{\text c}$. The region
of $\phi$ within $\pi/2 <\phi \leq \pi$ is the interesting region, 
for which we plot the polymer configurations.

To conclude: the ODS configuration of the chain resembles a rigid rod
below a critical value Wi$_{\text c}$ of the Weissenberg number. Above
this critical value, {\it even though the length of the chain is only
  about half as that of the persistence length}, it breaks the
reversal symmetry, much like the classic case of a pitchfork
bifurcation. This leads to the development of a region around
$\theta=\pi/2$ and $\phi=3 \pi/4$, where the chain (Euler)
buckles. The buckling gives a large dip in the end-to-end distance. 

Finally we note that the critical Wi$_{\text c}$ depends on the length $N$ of 
the chain (roughly inversely proportional) and on $\nu$ (decreasing with $\nu$).
Converting it to a critical shear rate $\dot{\gamma}_c$ involves also $T^*$
[see Eq.~(\ref{e4})].

\section{Shapes of a tumbling semiflexible polymer:
  simulations \label{sec6}}

Our simulations have been performed using an efficient implementation
of semiflexible polymer dynamics \cite{leeuwen2} of the bead-spring
model \cite{leeuwen,leeuwen1}.
\begin{figure*}
\includegraphics[width=0.49\linewidth]{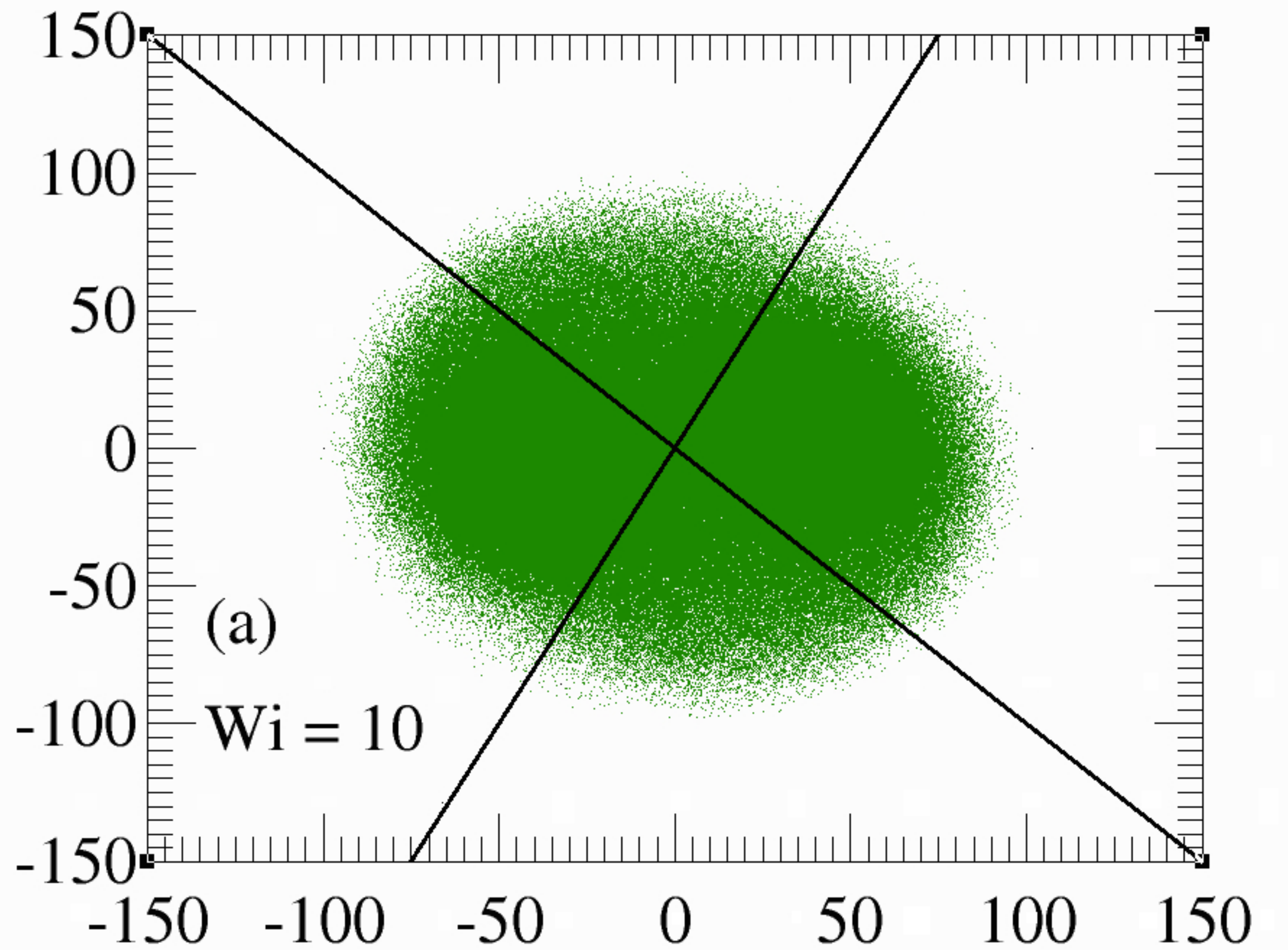}
\includegraphics[width=0.49\linewidth]{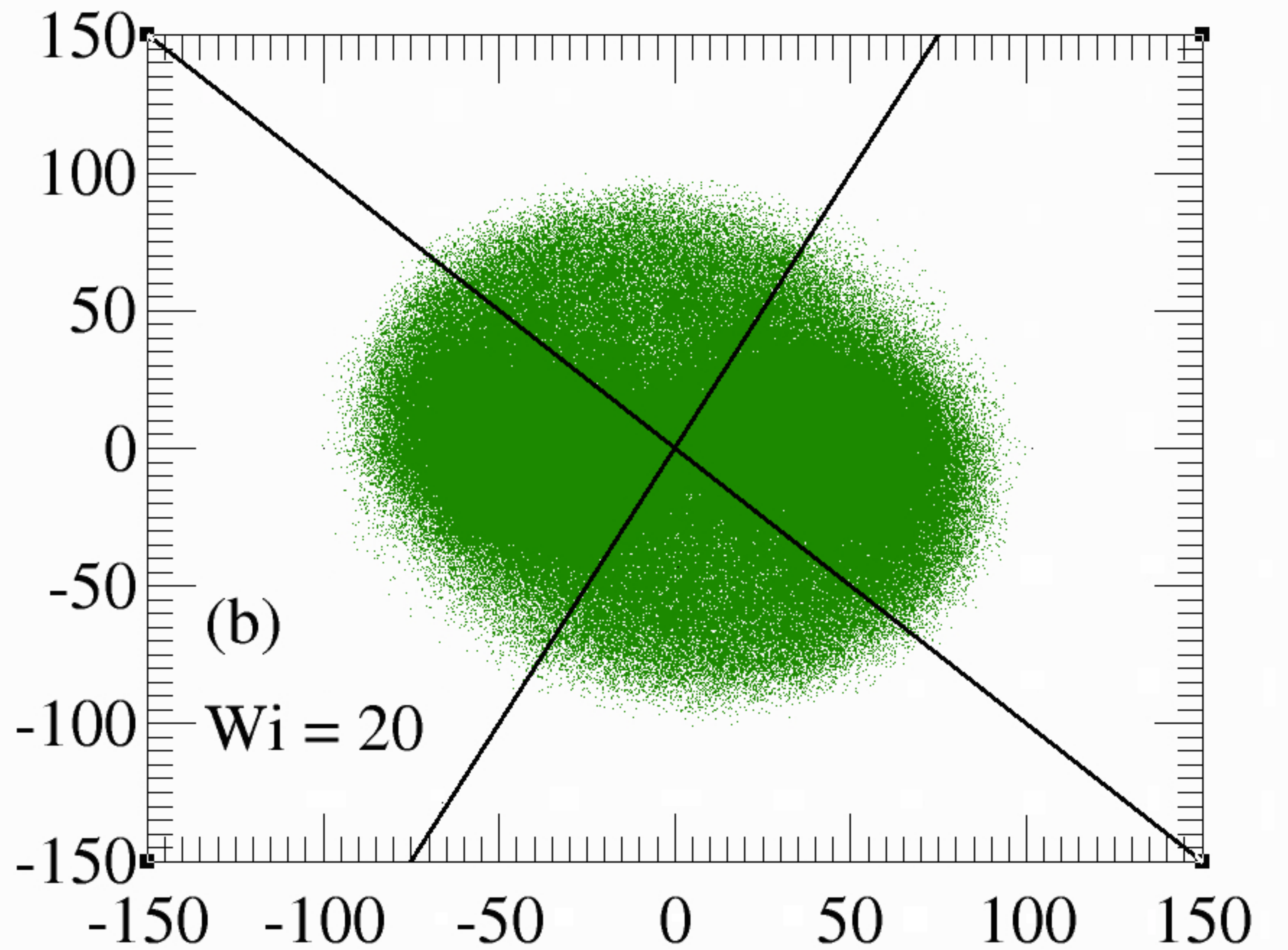}
\includegraphics[width=0.49\linewidth]{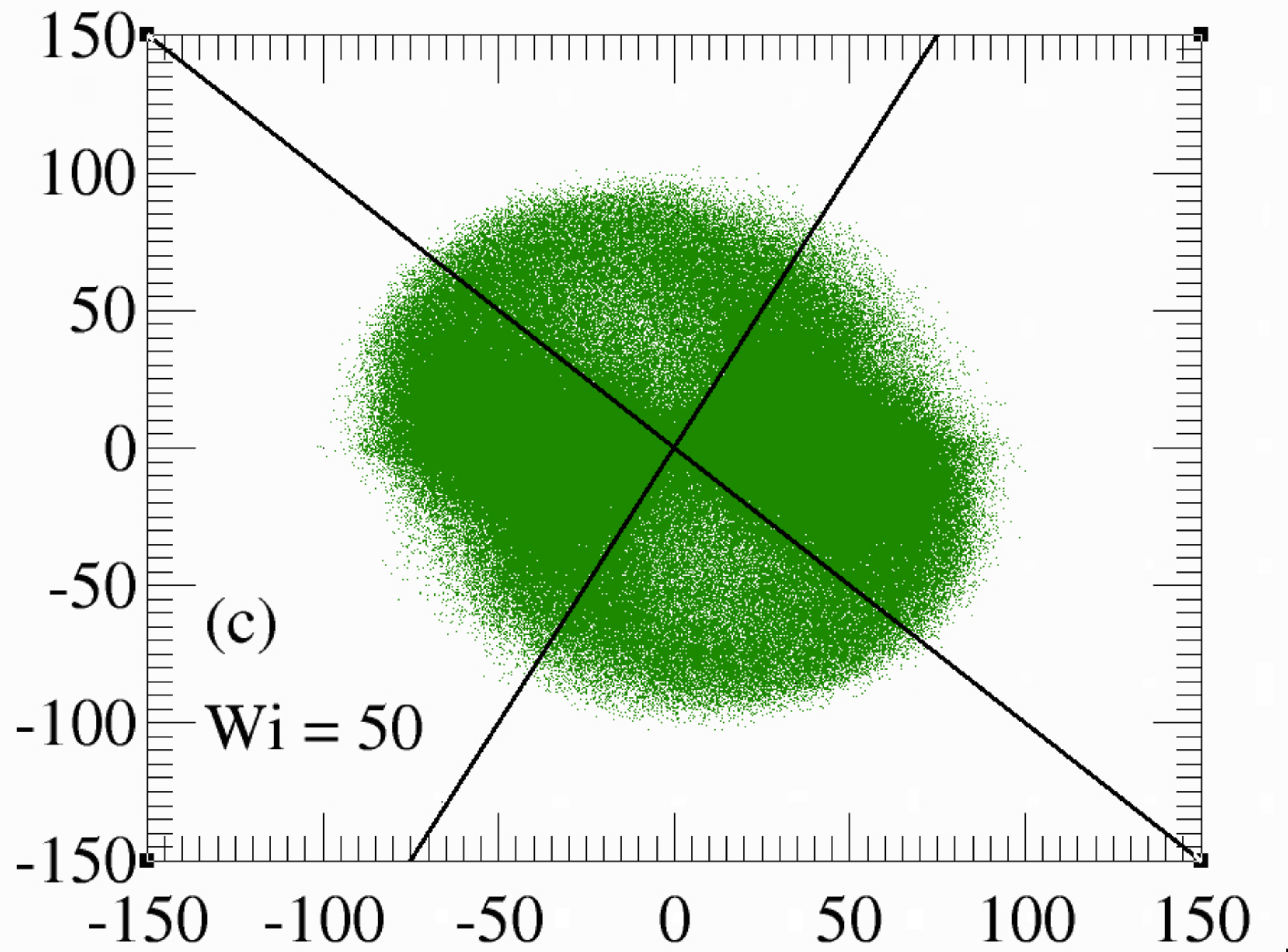}
\includegraphics[width=0.49\linewidth]{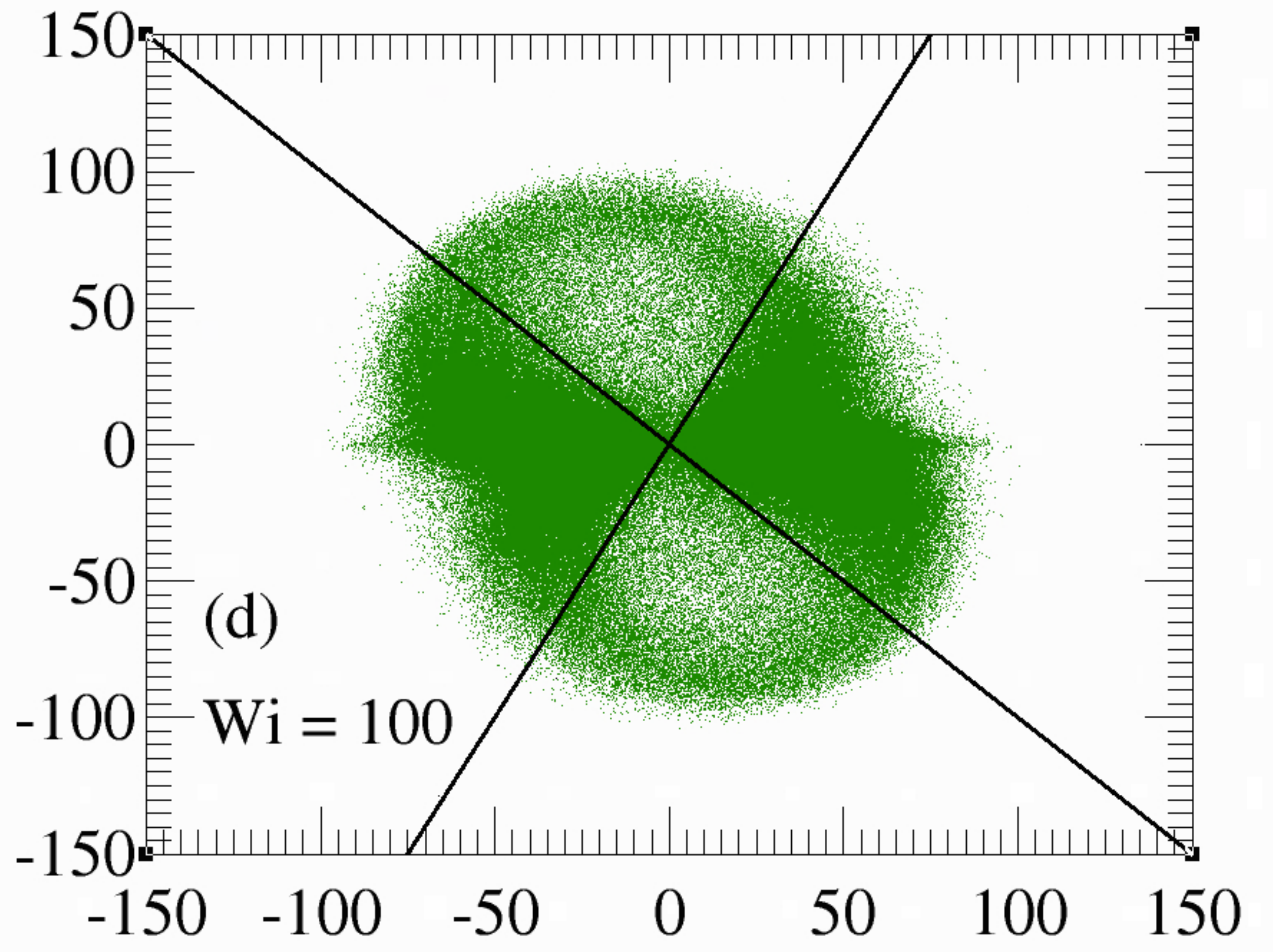}
\caption{(color online) Scatterplots of $R^s_2$ as a function of
  $\phi$ around $\theta=\pi/2$ at four values of Wi of length
  $N=63$. In between the solid lines, representing
  $0.35\pi<\phi<3\pi/4$, we see that the probability distribution
  $P(R^s_2)$ of $R^s_2$ changing with increasing Wi. We take these
  issues up in Fig. \ref{probdist}. See also the text for
  details. \label{scatter}}
\end{figure*}

Before we discuss the details of the simulation results, we make the
readers aware of the differences between the ODS in Sec. \ref{sec5}
and the simulations. Thermal noise plays no role in the ODS while in
simulations it does. This implies that although for Wi $<$ Wi$_{\text
  c}$ the amplitude of $R^s_2$ is identically zero in the ODS, we
should not expect to find the same in simulations at low Wi-values,
since in simulations the second Rouse mode will always be kicked up by
noise. This calls into question the relevance of the ODS for
simulations --- in particular, whether the tumbling of the chain is
sufficiently slow such that the simulation can explore the
neighborhood of the ODS, and thereby follow the characteristics of the
ODS. In view of lack of clarity for an answer to this question we used
the ODS as a guide for the simulations: to be more precise, we
focused on the values of Wi in the range 10 $\le$ Wi $\le$ 100 for
$N=63$ and sampled the probability distribution of $R^s_2$ as function
of the orientation.

In simulations for dsDNA of length $N=63$ we recorded 16 million
consecutive snapshots of the chain at regular intervals, ${\bf R}_1$
(for determining the orientation of the chain) and ${\bf R}_2$, at
equal intervals of time, for several values of Wi. The angles
$(\theta,\phi)$ for the chain's orientation are determined from the
values of ${\bf R}_1$. We then selected out the snapshots in this
slice $|\theta-\pi/2|=0.1$ radians, leaving us with 2-3 million
snapshots dependent on the value of Wi. In Fig. \ref{scatter} we show
the corresponding scatterplots of $R^s_2$ as a function of $\phi$ in
this slice at four values of Wi. In between the solid lines,
representing $0.35\pi<\phi<3\pi/4$, we see two lobes of empty regions
developing, signaling that the probability distribution $P(R^s_2)$ of
$R^s_2$ for these values of $\phi$ changes with increasing Wi. We note
that the values of $\phi$ corresponding to the two solid lines in
Fig. \ref{scatter} are chosen solely by visual inspection, and that
the locations of the empty lobes are shifted wrt the region in $\phi$,
for which the ODS exhibits Euler buckling.
\begin{figure}[h]
\includegraphics[width=0.49\linewidth]{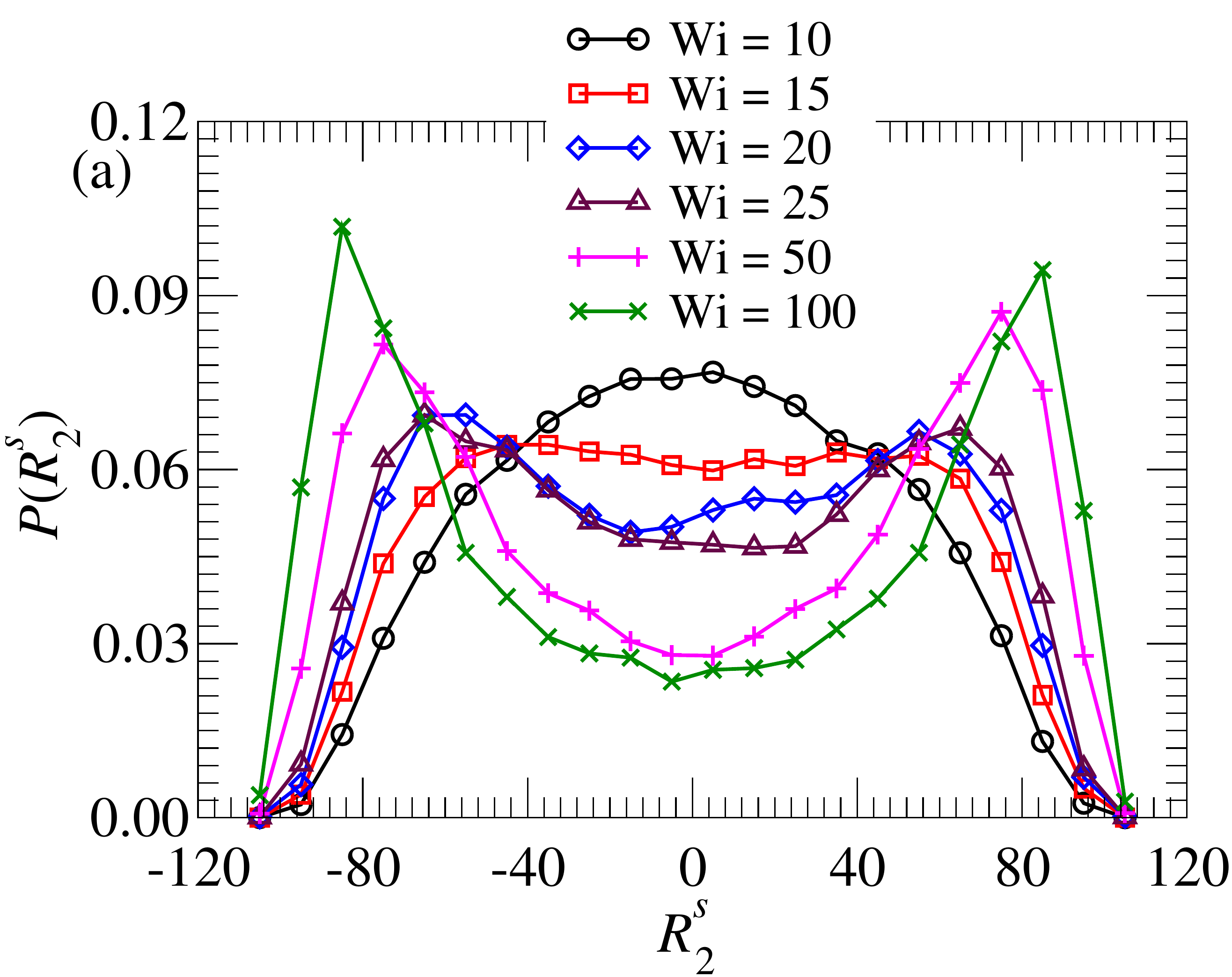}
\includegraphics[width=0.49\linewidth]{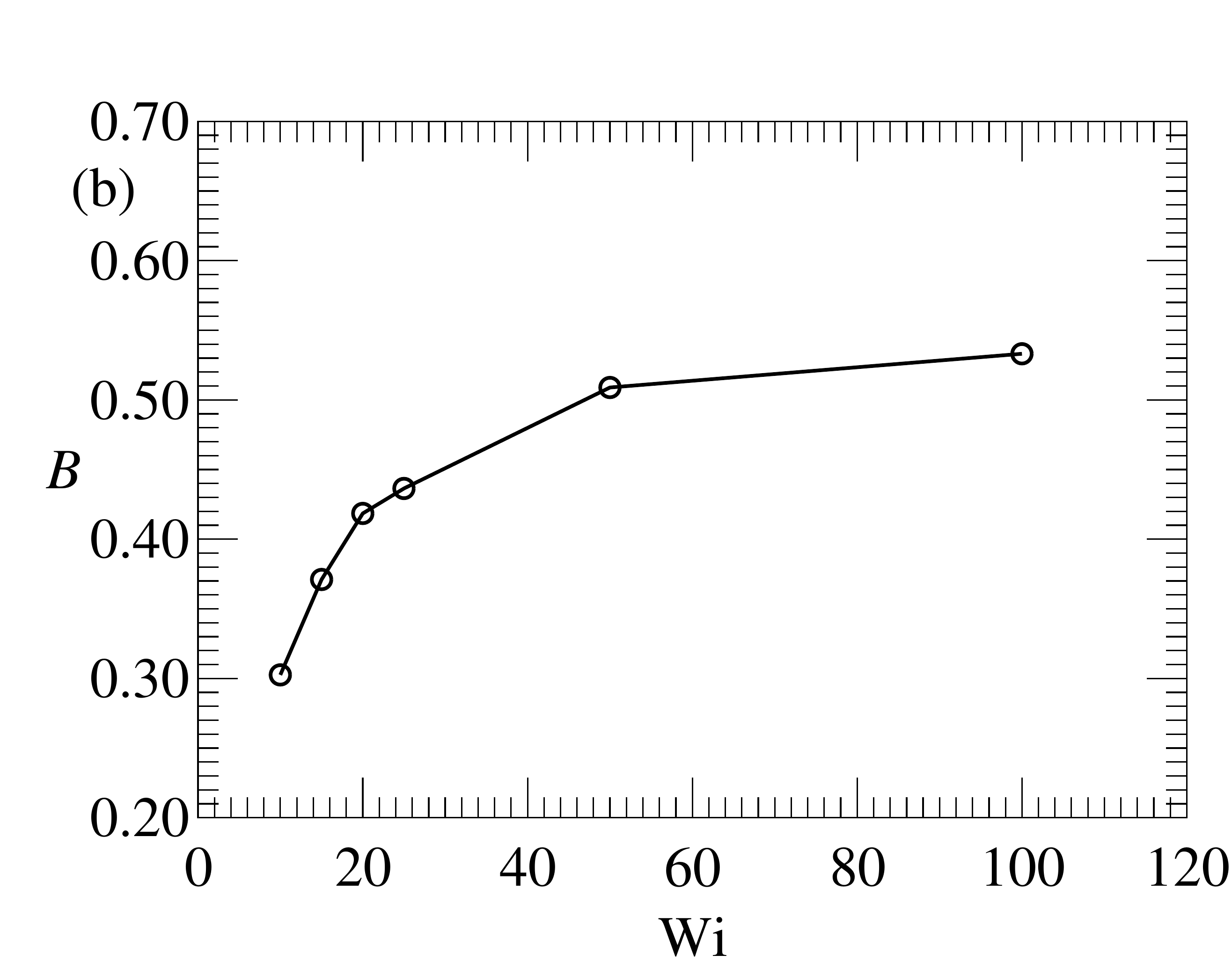}
\caption{(color online) (a) Probability distribution $P(R^s_2)$ of
  $R^s_2$ corresponding to $0.35\pi<\phi<3\pi/4$ in Fig. \ref{scatter}
  for $N=63$, showing that the unimodal distribution for Wi $=$ 10
  gradually transforming into a bimodal distribution symmetric around
  $R^s_2=0$ for higher Wi. (b) The corresponding Binder cumulant $B$,
  as defined in Eq. (\ref{bc}), which changes from $\approx0.3$ at Wi
  $=$ 10 to $\approx0.53$ at Wi $=$ 100. \label{probdist}}
\end{figure}

In order to further study the change in the probability distribution
$P(R^s_2)$ of $R^s_2$, we selected out the data points corresponding
to $0.35\pi<\phi<3\pi/4$ in Fig. \ref{scatter}, leaving us
50,000-100,000 data points. From them we constructed the probability
distribution $P(R^s_2)$. The distributions, corresponding to Wi $=$
10, 15, 20, 25, 50 and 100 are shown in Fig. \ref{probdist}. In
Fig. \ref{probdist}(a) we see that the unimodal distribution for Wi
$=$ 10 gradually transforms into a bimodal distribution symmetric
around $R^s_2=0$ for higher Wi. This development is the telltale sign
of symmetry breaking, which can also be tracked by the development of
the Binder cumulant $B$, defined as
\begin{eqnarray}
B=1-\frac{\langle(R^s_2)^4\rangle}{3\langle(R^s_2)^2\rangle},
\label{bc}
\end{eqnarray}
and shown in Fig.~\ref{probdist}(b). The Binder cumulant,
originally introduced to study symmetry breaking, attains the value
zero when the probability distribution is Gaussian, and reaches the
value $2/3$ when the symmetry is fully broken, changing the
probability distribution into a combination of two symmetric
$\delta$-peaks. For the data in Fig.~\ref{probdist}(a) we see that the
value of $B$ changes from $\approx0.3$ at Wi $=$ 10 to $\approx0.53$
at Wi $=$ 100.

\begin{figure}[h]
\includegraphics[width=0.49\linewidth]{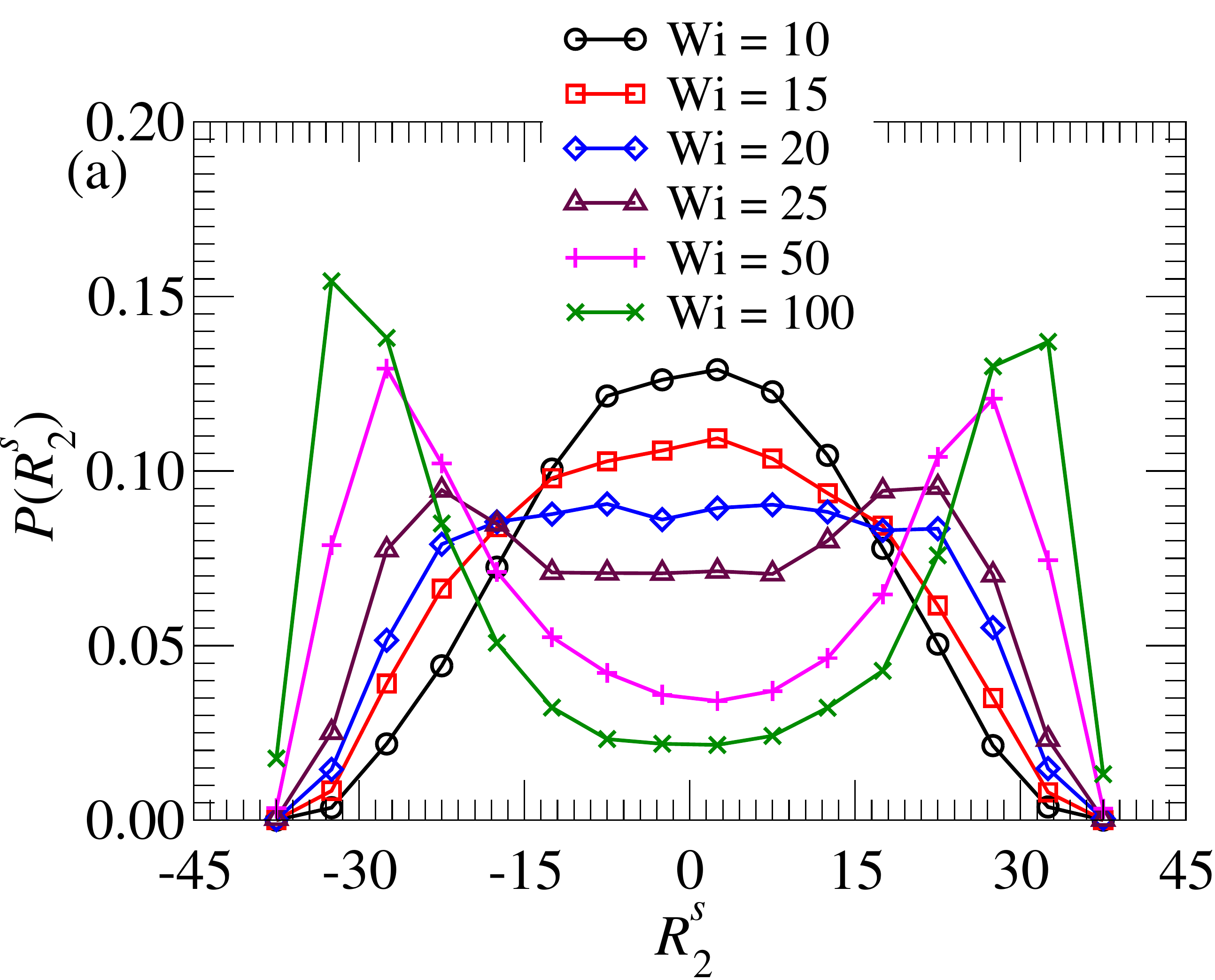}
\includegraphics[width=0.49\linewidth]{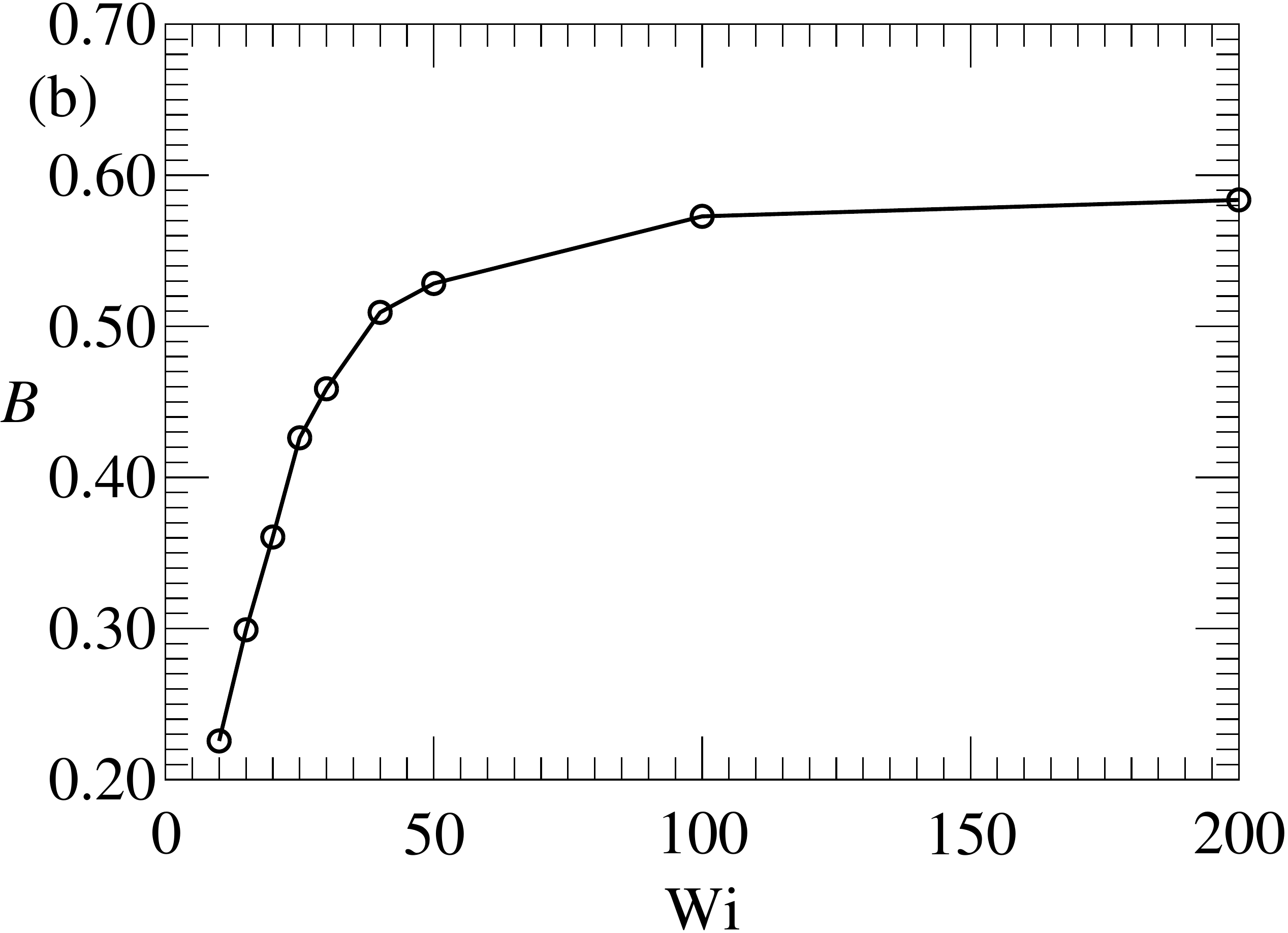}
\caption{(color online) (a) Probability distribution $P(R^s_2)$ of
  $R^s_2$ corresponding to $0.35\pi<\phi<3\pi/4$ in Fig. \ref{scatter}
  for $N=31$, showing that the unimodal distribution for Wi $=$ 10
  slowly transforming into a bimodal distribution symmetric around
  $R^s_2=0$ for higher Wi. (b) The corresponding Binder cumulant $B$,
  as defined in Eq. (\ref{bc}), which changes from $\approx0.23$ at Wi
  $=$ 10 to $\approx0.58$ at Wi $=$ 200. \label{probdist31}}
\end{figure}

The symmetry breaking is certainly not confined to
$N=63$. The same analysis on the simulation data (again, all data
points within $|\theta-\pi/2|=0.1$ and $0.35\pi<\phi<3\pi/4$, with the
corresponding figures, analogous to Fig.~\ref{probdist}, presented
in Fig.~\ref{probdist31}) reveals symmetry breaking taking place also
for $N=31$.

We note that the center of the region where the symmetry breaking
takes place is around the values of $\phi \approx 0.55 \pi$ as can be
observed from the scatterplots.  This is substantially different from
the value $\phi=0.75 \pi$ where the onset of buckling takes place in
the ODS. The chain tumbles in the direction from $\phi=\pi$ towards
$\phi=0$. So the buckling in the simulation lags behind with respect
to the ODS. This is likely the result of the slowness by which the
buckled state is formed and is broken down. Using Eq. (\ref{f1}) we
estimated the time $\Delta \tau$, needed to evolve from the
ground-state (in which the transverse ${\bf R}^2_2=0$) to 50\% of its
asymptotic value (the ODS), to be of the order $\Delta \tau \simeq
10^6$. This translates to a time $\Delta \tilde{\tau}\approx0.3$ [see
Eq. (\ref{e4})].  In order to put this estimate in perspective, we
compare it with the tumbling period $1/f$ of the rigid rod, which is
1.7 for Wi=20 according to Eq. (\ref{e8}). In other words, the chain
indeed travels a sizable fraction of the period in the building-up
phase of the buckling, the more so since it rotates faster for the
buckling orientations than in the position aligned with the flow.

Thus, to summarize this section: using the theoretical analysis of
symmetry breaking as a guide we have computed the probability
distribution $P(R^s_2)$ of $R^s_2$ by simulations of a tumbling dsDNA
segment of length $N=63$ and $N=31$. The simulation data has confirmed
that symmetry breaking takes place, showing up as the transition from
an unimodal probability distribution $P(R^s_2)$ of $R^s_2$ at Wi $=$
10 transforming into a bimodal distribution symmetric around
$R^s_2=0$, as well as the associated Binder cumulants.
\begin{figure}[h]
\includegraphics[width=0.49\linewidth]{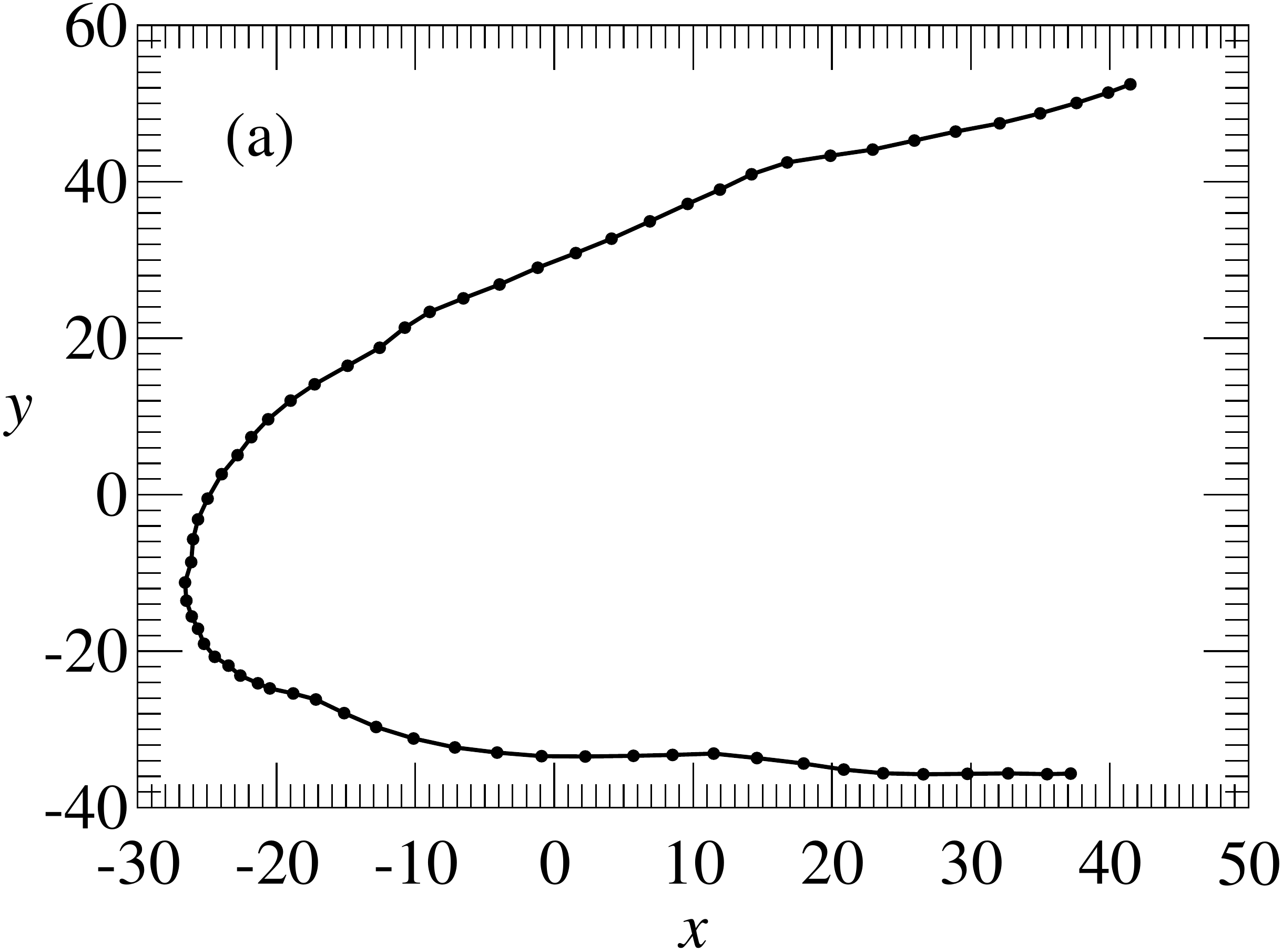}
\includegraphics[width=0.49\linewidth]{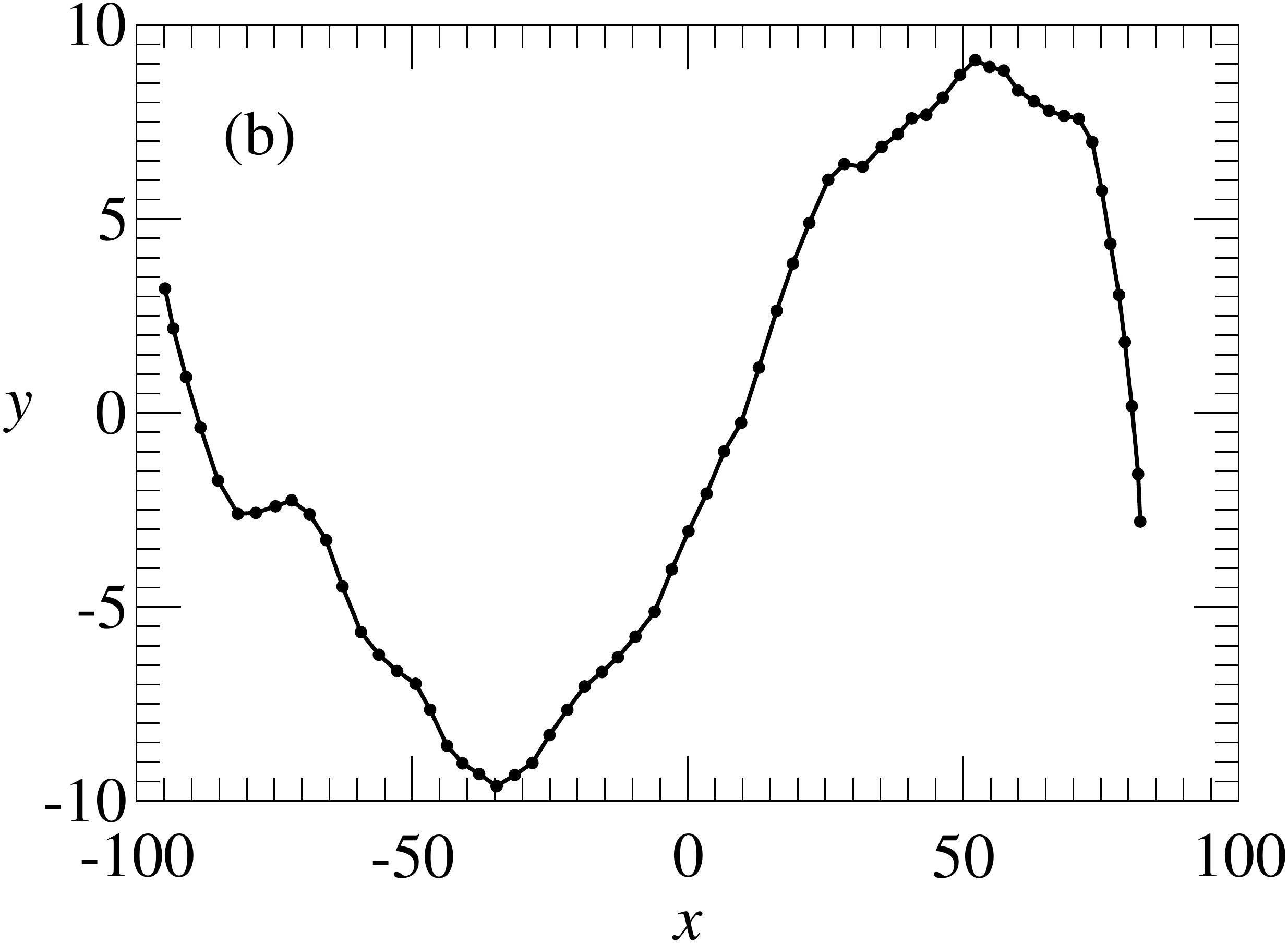}
\caption{Simulation snapshots a tumbling dsDNA chain of length $N=63$
  at Wi $=$ 100, projected on the $x$-$y$ plane: (a) U-shape (b)
  S-shape. \label{snapshots}}
\end{figure}

To supplement the above analysis of the simulation data we show, in
Fig.~\ref{snapshots}, two simulation snapshots of a tumbling dsDNA
chain of length $N=63$ at Wi $=$ 100, projected on the $x$-$y$ plane,
in order to showcase that, akin to the experimental snapshots shown
for f-actin in Ref. \cite{harasim}, shear can cause strong deformation
even for a chain that is shorter than its persistence length. A movie
of this tumbling chain (that includes both configurations of
Fig.~\ref{snapshots}) can be found in the ancillary files. In the 
movie the center-of-mass of the chain always remains at the origin 
of the co-ordinate system. The movie contains 3,000 snapshots, with
consecutive snapshots being $\Delta\tau=560$ apart in time. With
$\Delta\tau=1$ representing $0.16$ ps \cite{leeuwen2}, the full
duration of the movie spans $\approx2.7$ $\mu$s in real time.

\section{Conclusion\label{sec7}}

Our study focuses on fragments dsDNA, which are fairly extensible
semiflexible polymers. The extensibility of dsDNA implies parameters
in our Hamiltonian, which admit mode dynamics with a large time
step. The usual workhorse for theoretical studies is the inextensible
wormlike chain model for the Hamiltonian, the computer implementation
of which is confined to significantly smaller time steps.

Our simulations of a semiflexible polymer (dsDNA fragments smaller
than the persistence length) show that their tumbling frequency is
given, for the accessible range of Weissenberg numbers (Wi<2), by the
thin rigid-rod formula.  Deviations of the tumbling frequency from
this formula (Fig.~(\ref{raw})) occur at higher Weissenberg
numbers. It is theoretically interesting to speculate about the nature
of the deviations from the rigid-rod formula, also in view of the
observation that the accessible range of Weissenberg numbers is much
larger for stiffer and longer polymers, e.g. f-actin.  The Weissenberg
number for a polymer chain is a product of the shear rate $\dot\gamma$
and the rotational diffusion time-scale of a rigid rod of the same
length as the chain, i.e., $L$. Consequently, the Weissenberg number
$\propto \dot\gamma L^3$. One persistence length of f-actin is about
200 times longer than one persistence length of dsDNA. In units of
persistence length, for the same shear rate one thus reaches orders of
magnitude higher Weissenberg numbers for f-actin than for dsDNA.

In this respect we note that the Wi$^{2/3}$ law for rigid thin rods
originates from a singularity that develops in the probability
distribution for the orientation in the points $\theta=\pi/2$ and
$\phi=0$ or $\pi$ \cite{bloete}.  The reason is that a thin rigid rod
does not feel a torque from the shear in the aligned orientation and
only a fluctuation can pull the rod over this stagnation point. A
semiflexible polymer, however, always feels a torque due to
fluctuations of the other modes (either thermal or buckling), which
communicate with the orientation through the coupling force ${\bf
  H}_1$.  These fluctuations enable Jeffery-like orbits which are
characteristic for ellipsoids with a finite aspect ratio in the
moments of inertia \cite{burgers}.  The deviations from the thin rigid
rod formula that we see in Fig.~(\ref{raw}) do not substantiate the $f
\propto {\rm Wi}^{3/4}$ law reported for inextensible wormlike chains
\cite{lang}.

Using our Hamiltonian we have made a quantitative analysis of the
phenomenon of Euler buckling.  Fixing the orientation and searching
for the configuration which results by turning off the thermal noise,
yields the oriented deterministic state (ODS).  In the ODS we see a
sharply defined critical Wi$_c$ above which the buckling occurs. It is
a form of symmetry breaking through the occurrence of even modes in the
ODS above Wi$_c$.

In the simulations we observe correspondingly a transition in the
probability distribution for the even modes, in particular $R^s_2$,
changing gradually from a unimodal distribution to a bimodal
distribution. The simulations show that the formation of the buckled
state is a slow process. Therefore the orientation where the two peaks
in the bimodal are most significant, lags behind the orientation where
the ODS gives the maximum buckling.  The buckling is substantiated by
characteristic configurations and a movie of the tumbling process.

\end{document}